\documentclass[preprint,english,aps,pra,superscriptaddress,nofootinbib]{revtex4}%
\usepackage{amsfonts}
\usepackage[T1]{fontenc}
\usepackage[latin9]{inputenc}
\usepackage{amsmath}
\usepackage{amssymb}
\usepackage{babel}
\usepackage{graphicx}%
 
\begin{document}
\title{Relativistic Bohmian trajectories and Klein-Gordon currents for spin-0 particles}
\author{M.\ Alkhateeb}
\affiliation{Laboratoire de Physique Th\'eorique et Mod\'elisation, CNRS Unit\'e 8089, CY
Cergy Paris Universit\'e, 95302 Cergy-Pontoise cedex, France}
\author{A.\ Matzkin}
\affiliation{Laboratoire de Physique Th\'eorique et Mod\'elisation, CNRS Unit\'e 8089, CY
Cergy Paris Universit\'e, 95302 Cergy-Pontoise cedex, France}

\begin{abstract}
It is generally believed that the de Broglie-Bohm model does not admit a
particle interpretation for massive relativistic spin-0 particles, on the
basis that particle trajectories cannot be defined. We show this situation is
due to the fact that in the standard (canonical) representation of the Klein-Gordon
equation the wavefunction systematically contains superpositions of particle
and anti-particle contributions.\ We argue that by working in a
Foldy-Wouthuysen type representation uncoupling the particle from the
anti-particle evolutions, a positive conserved density for a particle and
associated density current can be defined.\ For the free Klein-Gordon equation
the velocity field obtained from this current density appears to be
well-behaved and sub-luminal in typical instances. As an illustration, Bohmian
trajectories for a spin-0 boson distribution are computed numerically for free propagation
in situations in which the standard velocity field would take arbitrarily high
positive and negative values.

\end{abstract}
\maketitle

\section{Introduction}

The de Broglie-Bohm interpretation of quantum mechanics \cite{dB,bohm} has
played and still plays an important role in our understanding of quantum
mechanics. Not that one should necessarily endorse the ontological package and mechanisms
put forward by the Bohmian model as describing the ``real'' behavior of quantum
systems.\ 
The strength of
the Bohmian model lies, in our view, elsewhere: by proposing an account of
dynamical processes for which the orthodox interpretation tells us we should
give up any attempt to explain them, the de Broglie-Bohm interpretation
improves our understanding and our intuition of quantum phenomena.

However, this situation holds for non-relativistic quantum mechanics based on
the Schrödinger equation. In the relativistic domain, the Bohmian model
suffers from several difficulties. In particular, it seems impossible, to
define trajectories for spin-0\ bosons described by the Klein-Gordon equation
(see Ch.\ 11 of Ref. \cite{undivided}). Obviously, one might expect that in
the high-energy regimes in which particles can be created or destroyed, the
usual Bohmian framework based on continuous trajectories would need to be
replaced by a different picture based on quantum field theory. 
But
it might appear surprising that even in low energy regimes, it is not possible
to give a de Broglie-Bohm trajectory description of a spin-0 particle propagating in free space.

The aim of this work is to examine the underlying reasons for the breakdown of
the Bohmian model for systems obeying the Klein Gordon equation, point out
how the problems can be repaired, and effectively introduce Bohmian trajectories from a
newly defined Klein-Gordon current.\ In a nutshell, our view is that in
relativistic quantum mechanics, a quantum state intrinsically superposes
particles and anti-particles. In particular, causality only holds if the time
evolution operator includes the propagator for the particle and anti-particle
sectors. So a generic state describing a particle (with a positive density
everywhere) will develop anti-particle components, and the density will become
negative in some spatial regions. This leads to superluminal velocities and closed loops in space-time \cite{dewd}, a feature that is deemed "inconsistent" (see
Sec.\ 10.4\ of Ref.\ \cite{undivided}, or \cite{holland-PR}) with the guidance
formula at the basis of the Bohmian model.\ This is why in the past some
radical solutions have been proposed, such as restricting the set of
admissible states to those that lead to positive only densities \cite{kypra},
or employing the absolute value of the density in order to define trajectories
\cite{nikolic08}. The first solution does not work, as remarked by Bell (see
footnote in \cite{kypra}), while the second one still leads to superluminal
velocities for the Bohmian particle. Other approaches have also been suggested
\cite{vigier-H,dewdney2000,holland2019,vollick2021}.

The solution we will advocate will be to work in a representation that
uncouples the particle and anti-particle sectors.\ In momentum space, this is
well-known to be possible through Foldy-Wouthuysen-type transformations
\cite{FW,case,FV} (see also \cite{silenko} for a recent review). Then a new
density needs to be defined in this representation. A locally conserved
current different from the canonical one can indeed be defined in a given reference
frame, though it is known that such currents do not transform as a 4-vector
\cite{hoffmann}.

We will begin (Sec.\ 2) by recalling how the Bohmian velocity field naturally
appears in non-relativistic quantum mechanics, even without imposing the
guidance formula. We will then examine why the same procedure does not work
with the Klein-Gordon equation in its canonical form. In Sec.\ 3 we will introduce the
representation separating the particle and anti-particle sectors and introduce
the density and corresponding current defined directly in that representation.
We will then define de Broglie-Bohm trajectories in that representation and compute trajectories
for the free Klein-Gordon equation (Sec.\ 4). We close with a few concluding remarks (Sec.\ 5).

\section{Density current and particle velocity}

\subsection{Non-relativistic guidance equation}

In the usual non-relativistic de Broglie-Bohm approach, the particle velocity
$v_{j}(x,t)$ is obtained from the standard Schrödinger current $j_{NR}(x,t)$
through%
\begin{equation}
v_{j}(x,t)=\frac{j_{NR}(x,t)}{\rho_{NR}(x,t)} \label{vNR}%
\end{equation}
where%
\begin{equation}
\rho_{NR}(x,t)=\left\vert \left\langle x\right\vert \left.  \psi\right\rangle
\right\vert ^{2} \label{rhoNR}%
\end{equation}
is the probability density, whose conservation takes the form%
\begin{equation}
\frac{\partial\rho_{NR}}{\partial t}+\partial_{x}j_{NR}(x,t)=0
\end{equation}
(to simplify the notation, we will assume a free Hamiltonian and stay in one spatial dimension). Recall that $v_{j}(x,t)$ defines a velocity
field, and that the Bohmian particle follows the streamlines of the current:
the trajectory $x(t)$ is obtained by integrating the guidance equation
\begin{equation}
\frac{dx}{dt}=v_{j}\left(  x(t),t\right)
\end{equation}
with an initial condition $x(t_{0})=x_{0}$.

The current can be written as the symmetrized combination of the momentum
operator $P$ and a spatial projection:
\begin{equation}
j_{NR}(x,t)=\frac{1}{2m}\left(  \left\langle \psi\right\vert P\left\vert
x\right\rangle \left\langle x\right\vert \left.  \psi\right\rangle
+\left\langle \psi\right\vert \left.  x\right\rangle \left\langle x\right\vert
P\left\vert \psi\right\rangle \right)  , \label{jschro}%
\end{equation}
and the average of the current density is easily seen to yield
\begin{equation}
\left\langle j_{NR}(t)\right\rangle =\int dxj_{NR}(x,t)=\left\langle
\psi\right\vert \frac{P}{m}\left\vert \psi\right\rangle \equiv\left\langle
\psi\right\vert V\left\vert \psi\right\rangle \label{avschro}%
\end{equation}
where $V\equiv P/m$ can be seen as a velocity operator.\ Hence the average
Schrödinger current gives the average velocity of the system.

It is also straightforward to write Eq. (\ref{jschro}) as
\begin{equation}
j_{NR}(x,t)=\left\vert \left\langle x\right\vert \left.  \psi\right\rangle
\right\vert ^{2}\left[  \frac{1}{2}\left(  \frac{\left\langle \psi\right\vert
\frac{P}{m}\left\vert x\right\rangle }{\left\langle \psi\right\vert \left.
x\right\rangle }+\frac{\left\langle x\right\vert \frac{P}{m}\left\vert
\psi\right\rangle }{\left\langle x\right\vert \left.  \psi\right\rangle
}\right)  \right]  . \label{rhov}%
\end{equation}
Comparing with Eqs. (\ref{vNR}) and (\ref{rhoNR}), we see that the term
between $[...]$ is precisely the velocity (\ref{vNR}) that we can rewrite as
\begin{equation}
v_{j}(x,t)=\operatorname{Re}\frac{\left\langle x\right\vert V\left\vert
\psi\right\rangle }{\left\langle x\right\vert \left.  \psi\right\rangle }.
\end{equation}
The expression on the right handside is sometimes known as the weak value of the
velocity operator \footnote{This expression was first obtained in Ref.
\cite{leavens}; see \cite{WVreview} for a brief review on weak values, including
a discussion on the current density.}. The important point is that the velocity
of the particle in the non-relativistic de Broglie-Bohm approach is naturally
defined from the wavefunction and the velocity operator.

\subsection{Klein-Gordon equation}

\subsubsection{Current and density in the canonical formulation}

The Klein-Gordon equation in its usual \textquotedblleft
canonical\textquotedblright\ form is \cite{greiner}%
\begin{equation}
(i\hbar\partial_{t})^{2}\psi=\left(  P^{2}c^{2}+m^{2}c^{4}\right)
\psi\label{kg-f}%
\end{equation}
and the associated density reads%
\begin{equation}
\rho(x,t)=\frac{i\hbar}{2mc^{2}}\left(  \psi^{\ast}(x,t)\partial_{t}%
\psi(x,t)-\psi(x,t)\partial_{t}\psi^{\ast}(x,t)\right)  ; \label{kgc-rho}%
\end{equation}
$m$ is the rest mass and $c$ the light velocity and the $\partial_{t}$ term
entering the definition of the scalar product is due to the presence of a
second order time derivative in Eq. (\ref{kg-f}). The density is not positive
definite (this will become clearer below), and $\rho$ is therefore interpreted
as a charge density. Note that in the free case plane waves with both positive
and negative energies are solutions of Eq. (\ref{kg-f}), associated with
particles and anti-particles respectively \cite{greiner}.

The corresponding conserved current obeying $\partial_{t}\rho+\partial_{x}j=0$
can be written in the same form as the Schrödinger current (\ref{jschro}),
namely%
\begin{equation}
j(x,t)=\frac{1}{2m}\left(  \left\langle \psi\right\vert P\left\vert
x\right\rangle \left\langle x\right\vert \left.  \psi\right\rangle
+\left\langle \psi\right\vert \left.  x\right\rangle \left\langle x\right\vert
P\left\vert \psi\right\rangle \right)  . \label{kgc-j}%
\end{equation}
We therefore also have as for the Schr\"{o}dinger current average%
\begin{equation}
\left\langle j(t)\right\rangle =\left\langle \psi\right\vert \frac{P}%
{m}\left\vert \psi\right\rangle . \label{kgc-avj}%
\end{equation}
But in a relativistic context, $p/m$ is not the classical velocity of a
particle (it is not even bounded); the classical expression for the
relativistic velocity is%
\begin{equation}
v_{cl}=\frac{pc^{2}}{E_{p}}=\frac{p}{\gamma m}, \label{class}%
\end{equation}
where $E_{p}\equiv\sqrt{p^{2}c^{2}+m^{2}c^{4}}$. Hence there is no analog for
the right hand-side of Eq. (\ref{avschro}). Moreover the current cannot take
the form (\ref{rhov}), as $\left\vert \left\langle x\right\vert \left.
\psi\right\rangle \right\vert ^{2}$ does not represent the conserved density.

\subsubsection{Klein-Gordon equation in Hamilton form}

As is well-known \cite{greiner}, the Klein-Gordon (KG) equation can be written
in a form involving a first order time derivative,%
\begin{equation}
i\hbar\partial_{t}\Psi=\left[  \left(  \tau_{3}+i\tau_{2}\right)  \frac{P^{2}%
}{2m}+\tau_{3}mc^{2}\right]  \Psi\label{H-form}%
\end{equation}
where $\tau_{i}$ are the usual Pauli matrices (generally denoted $\sigma_{i}$
in other contexts). The wave-function
\begin{equation}
\Psi=\left(
\begin{array}
[c]{c}%
\varphi\\
\chi
\end{array}
\right)
\end{equation}
is now a 2-dimensional vector whose components are related to the solutions of
the canonical KG\ equation (\ref{kg-f}) by%
\begin{align}
\psi &  =\varphi+\chi\\
i\hbar\partial_{t}\psi &  =mc^{2}\left(  \varphi
-\chi\right)  .
\end{align}
The positive and negative energy plane-wave solutions of the canonical KG
equation become%
\begin{align}
\Psi^{+}  &  =\frac{1}{2\sqrt{mc^{2}E_{p}}}\left(
\begin{array}
[c]{c}%
mc^{2}+E_{p}\\
mc^{2}-E_{p}%
\end{array}
\right)  e^{-i\left(  E_{p}t-px\right)  /\hbar}\label{solpos}\\
\Psi^{-}  &  =\frac{1}{2\sqrt{mc^{2}E_{p}}}\left(
\begin{array}
[c]{c}%
mc^{2}-E_{p}\\
mc^{2}+E_{p}%
\end{array}
\right)  e^{i\left(  E_{p}t-px\right)  /\hbar}. \label{solneg}%
\end{align}

The density (\ref{kgc-rho}) and current (\ref{kgc-j}) take the form%
\begin{align}
\rho &  =\left\langle \Psi\right\vert \tau_{3}\left\vert x\right\rangle
\left\langle x\right\vert \left.  \Psi\right\rangle =\left\vert \varphi
(x)\right\vert ^{2}-\left\vert \chi(x)\right\vert ^{2}\label{density-H}\\
j  &  =\frac{1}{2}\left(  \left\langle \Psi\right\vert \tau_{3}\left\vert
x\right\rangle \left\langle x\right\vert \left(  \tau_{3}+i\tau_{2}\right)
\frac{P}{m}\left\vert \Psi\right\rangle +\left\langle \Psi\right\vert \tau
_{3}\left(  \tau_{3}+i\tau_{2}\right)  \frac{P}{m}\left\vert x\right\rangle
\left\langle x\right\vert \left.  \Psi\right\rangle \right)  \label{current-H}%
\end{align}
Note that the right hand-side of Eq. (\ref{density-H}) indicates that
$\chi(x)$ contributes to a negative charge and can thus be seen as an
anti-particle contribution. The $\tau_{3}$ matrix in the dual vector is the
signature of the pseudo-Hermitian character of the KG formalism \ \cite{mosta}%
, the (non-positively defined) inner product being given by $\left\langle
\Psi_{1}\right\vert \left.  \Psi_{2}\right\rangle _{KG}=\left\langle \Psi
_{1}\right\vert \tau_{3}\left\vert \Psi_{2}\right\rangle .$ The momentum
operator is multiplied by $\tau_{3}+i\tau_{2},$ changing the sign of the
anti-particle contribution.

\begin{figure}[t]
	\centering \includegraphics[width=12cm]{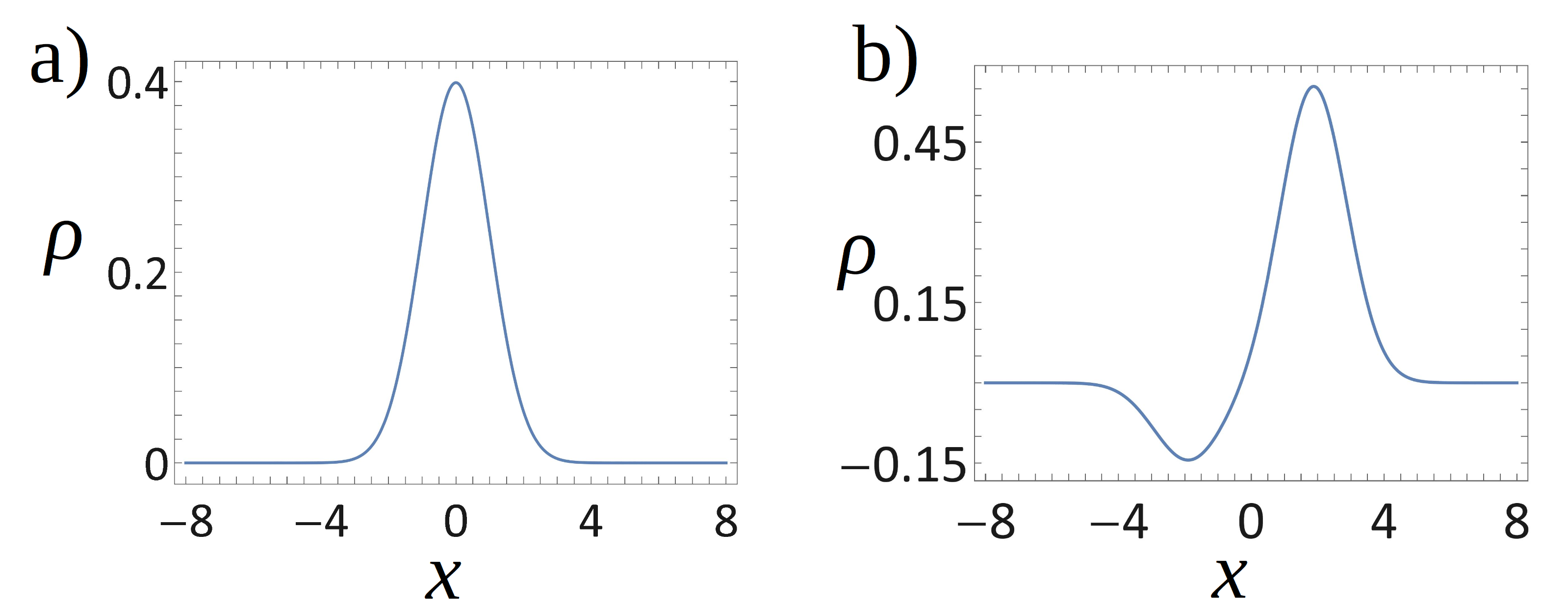}\caption{The canonical Klein-Gordon density $\rho(x,t)$ of
		an ultra-relativistic spin-0 boson (a) at $t=0$ and (b) at $t=2$ (natural units $\hbar=c=m=1$ are used). At $t=0$
		the distribution is a Gaussian (with mean momentum $p_0=3$, mean position $x_0=0$ and unit spatial width) that is everywhere positive, but the density becomes negative in some regions as the 
		wavefunction evolves.}%
	\label{fig-ex-vel}%
\end{figure}

\subsubsection{Standard Bohmian particle velocity}

The Hamiltonian formulation of the KG equation -- a system of two coupled
linear equations -- is handy in showing that a normalized initial state
$\Psi(t_{0},x)=\left(
\begin{array}
[c]{c}%
\varphi(t_{0},x)\\
0
\end{array}
\right)  $ with positive charge $\rho=\left\vert \varphi(t_{0},x)\right\vert
^{2}$ everywhere at $t=t_{0}$ will become at an evolved time $\Psi
(t,x)=\left(
\begin{array}
[c]{c}%
\varphi(t,x)\\
\chi(t,x)
\end{array}
\right)  ,$ so that typically $\rho(t,x)$ \ will become negative in some
regions. Attempting to
define a Bohmian velocity through%
\begin{equation}
v(t,x)\overset{?}{=}\frac{j(t,x)}{\rho(t,x)} \label{vel-c}%
\end{equation}
will lead to regions of infinitely high velocities.\ This happens in the
vicinity of a point for which $\rho(t,x)=0,$ but this does not necessarily
imply that $j(x)\rightarrow0$. Indeed, according to Eq. (\ref{density-H}), an
equal amount of positive and negative charge yields a vanishing density, i.e.
$\left\vert \varphi(x)\right\vert \rightarrow\left\vert \chi(x)\right\vert
\Longrightarrow\rho(x)\rightarrow0$.

\begin{figure}[t]
	\centering \includegraphics[width=9cm]{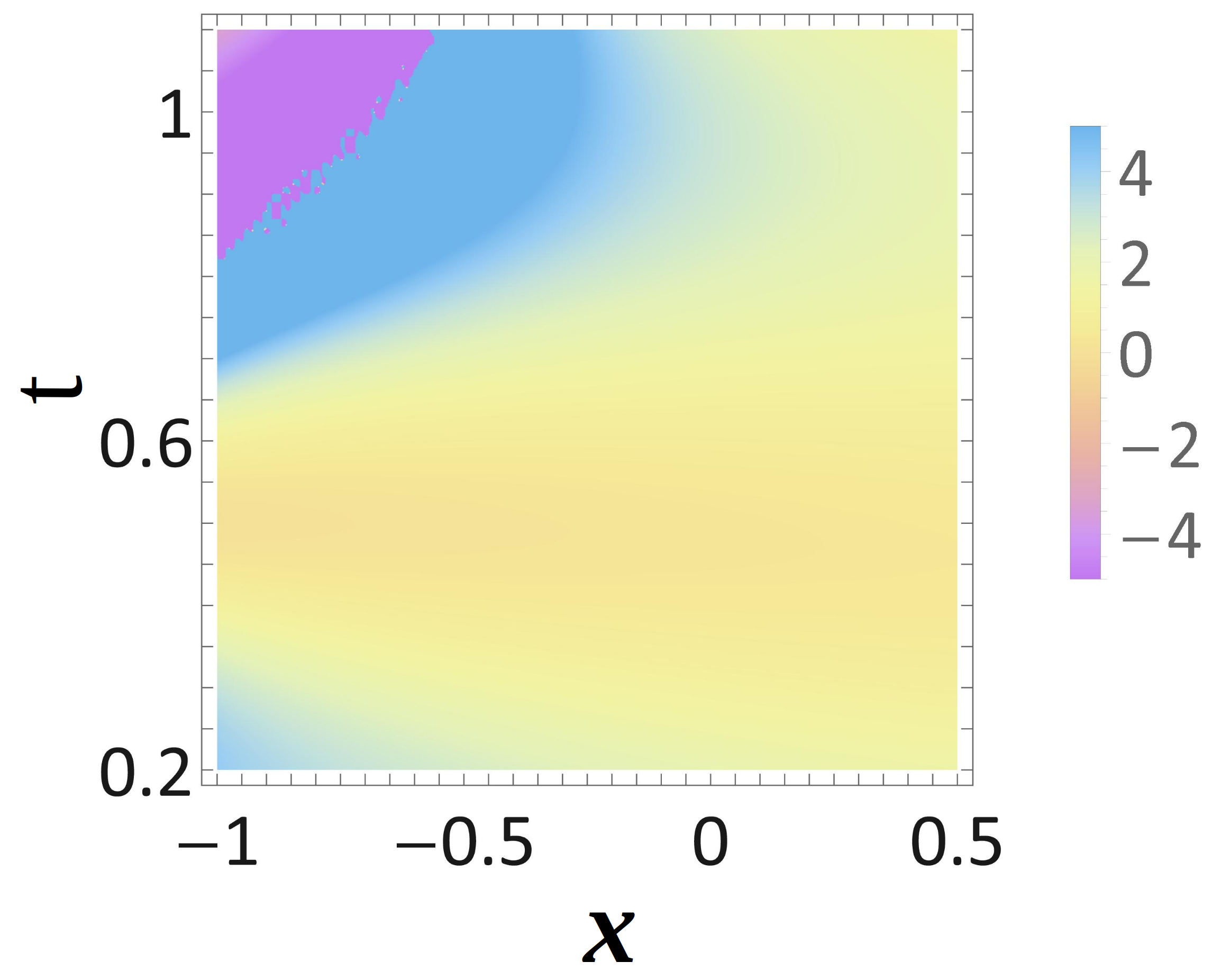}\caption{The velocity field defined by Eq. (\ref{vel-c}) in a given space-time region
		for the initial density shown in Fig. \ref{fig-ex-vel}(a) (natural units are used). The velocity field is superluminal in some spatial regions.}%
	\label{fig-vel-map-c}%
\end{figure}

An illustration is given in Fig. \ref{fig-ex-vel}: an initial positive charged
Gaussian (Fig. \ref{fig-ex-vel}(a)) quickly develops regions of negative
charge density (Fig. \ref{fig-ex-vel}(b)). The velocity field, as defined by
Eq. (\ref{vel-c}) is shown in Fig. \ref{fig-vel-map-c}. Regions of
superluminal velocities are readily visible, as expected when $\rho(t,x)$ goes
through 0 and changes sign. The wavefunctions were computed semi-analytically from the
wavepacket expansion of the time-evolved Gaussian in the Hamiltonian representation (see \cite{mak} for details).

While it is generally acknowledged that having a superluminal motion ruins a
particle trajectory interpretation of spin-0\ bosons \cite{undivided,dewd,holland-book}, some
authors, beginning with de Broglie \cite{debroglie-KGsup} (see also
\cite{nikolic04}) have asserted that this type of motion is not a problem as
long as it does not have experimental consequences (that is, this type of
behavior for individual particles is washed out by the intrinsically
statistical nature of the quantum formalism). One of us has argued elsewhere
\cite{pitfall,billiard} on general grounds why, if the de Broglie-Bohm
interpretation is taken as a realist construal of quantum phenomena, such
arguments should be rejected, as they undermine realism (since any fundamental
dynamical law can be postulated as long as the statistical predictions are
recovered) and employ the same type of strategy as the Copenhagen
interpretation: what matters in the predictive agreement with experiments, and an experiment 
that would specifically measure velocities will never detect superluminal aspects \footnote{De Broglie wittingly notes the similarity between his reasoning and Bohr's argumentation, see p. 135 in \cite{debroglie-KGsup}.}.
 For this reason, in our view a particle interpretation for spin-0 bosons hinges on the existence of well-behaved trajectories.

\section{Separating particles from anti-particles}

\subsection{Pseudo-unitary transformations}

Representations decoupling the positive and energy components were looked for
soon after the classic Newton-Wigner work \cite{NW} on the correct form of the
position operator in relativistic quantum mechanics. This was first worked out
in the context of the Dirac equation by Foldy and Wouthuysen \cite{FW}, and
then generalized to the Klein-Gordon case and particles of arbitrary spin
\cite{case,FV}. It was indeed realized \cite{costella,silenko} that the
representation in which $X$ is the position operator is not the standard Dirac
one, nor in the KG case the Hamilton formulation given by Eq. (\ref{H-form}),
but a representation in which the particle and anti-particle components are
not coupled.

The unitary transformation (or rather, pseudo-unitary in the KG case) depends
on the specific Hamiltonian of the problem. It is known in explicit form in a
few special cases, including the free Hamiltonian. For the free particle KG
Hamiltonian (\ref{H-form}), a well-known operator \cite{greiner} that
separates particles from anti-particles is%
\begin{equation}
U=\frac{\left(  mc^{2}+E_{p}\right)  -\tau_{1}\left(  mc^{2}-E_{p}\right)
}{\sqrt{4mc^{2}E_{p}}}.
\label{FW-mat}
\end{equation}
Note that $U$ is pseudo-unitary in the sense that $U^{\dagger}\neq U^{-1}$ but
$\tau_{3}U^{\dagger}\tau_{3}=U^{-1}$. The positive and negative energy
solutions (\ref{solpos}) and (\ref{solneg}) become%
\begin{align}
\Phi^{+}(p,t)  &  =U\Psi^{+}=\left(
\begin{array}
[c]{c}%
1\\
0
\end{array}
\right)  e^{-i\left(  E_{p}t-px\right)  /\hbar}\\
\Phi^{-}(p,t)  &  =U\Psi^{-}=\left(
\begin{array}
[c]{c}%
0\\
1
\end{array}
\right)  e^{i\left(  E_{p}t-px\right)  /\hbar}.
\end{align}
$\Phi^{+}(p,t)$ defines indeed a state with pure particle contribution.\ The
KG equation (\ref{H-form}) becomes%
\begin{equation}
i\hbar\partial_{t}\Phi=H_{\Phi}\Phi\label{KG-FW}%
\end{equation}
with%
\begin{equation}
H_{\Phi}=UHU^{-1}=\tau_{3}\sqrt{p^{2}c^{2}+m^{2}c^{4}}, \label{hamFW}%
\end{equation}
Since the Hamiltonian in this representation -- which, contrary to Eq.
(\ref{H-form}), has the same form as the classical relativistic Hamiltonian --
is diagonal, the evolution operator will conserve the pure particle character
of an initial particle wavepacket, without the appearance of any anti-particle contribution.

\subsection{Density and current}

The standard KG density and current, given by Eqs. (\ref{density-H}) and
(\ref{current-H}) can be written in the uncoupled representation, but this
will not change the property or the values taken by these quantities. We
should instead define a new density and a new current density directly from
the uncoupled wavefunction.

Let us assume we have a particle wavepacket, taken as a linear superposition%
\begin{equation}
\Phi(x,t)=\int dpg(p)\Phi^{+}(p,t),
\end{equation}
that we can rewrite more simply as
\begin{equation}
\phi(x,t)=\int dpg(p)e^{-i\left(  E_{p}t-px\right)  /\hbar}%
\end{equation}
with $\Phi(x,t)\equiv\left(  \phi(x,t),0\right)  $. $\phi(x,t)$ obeys the
upper line of Eqs. (\ref{KG-FW})-(\ref{hamFW}), an equation that is sometimes
known as the Salpeter or relativistic Schrödinger equation \cite{kowalski}.

Let us define the density $\varrho(x,t)$ by
\begin{equation}
\varrho(x,t)=\left\vert \phi(x,t)\right\vert ^{2}=\int dkdpg^{\ast}%
(k)g(p)e^{-i\left(  E_{p}-E_{k}\right)  t/\hbar}e^{-i\left(  k-p\right)
x/\hbar}. \label{RHO}%
\end{equation}
$\varrho(x,t)$ is of course positive by definition. It can then be shown
\cite{kowalski} that the quantity%
\begin{equation}
J(x,t)=c^{2}\int dkdp\frac{p+k}{E_{p}+E_{k}}g^{\ast}(k)g(p)e^{-i\left(
E_{p}-E_{k}\right)  t/\hbar}e^{-i\left(  k-p\right)  x/\hbar} \label{J}%
\end{equation}
defines a current density obeying the continuity equation $\partial_{t}%
\varrho+\partial_{x}J=0$. By integrating over $x$, it is straightforward to
obtain%
\begin{equation}
\left\langle J(t)\right\rangle =\left\langle \phi\right\vert \frac{Pc^{2}%
}{E_{p}}\left\vert \phi\right\rangle . \label{avJ}%
\end{equation}
This expression is the average of an operator that is the quantized
counterpart of the classical relativistic velocity. This was not the case with
the average of the canonical KG current, given by Eq. (\ref{kgc-avj}),
although in the non-relativistic case, the average of the Schrödinger current
(\ref{avschro}) does match the expression for the average (non-relativistic) velocity.

\begin{figure}[t]
	\centering \includegraphics[width=12cm]{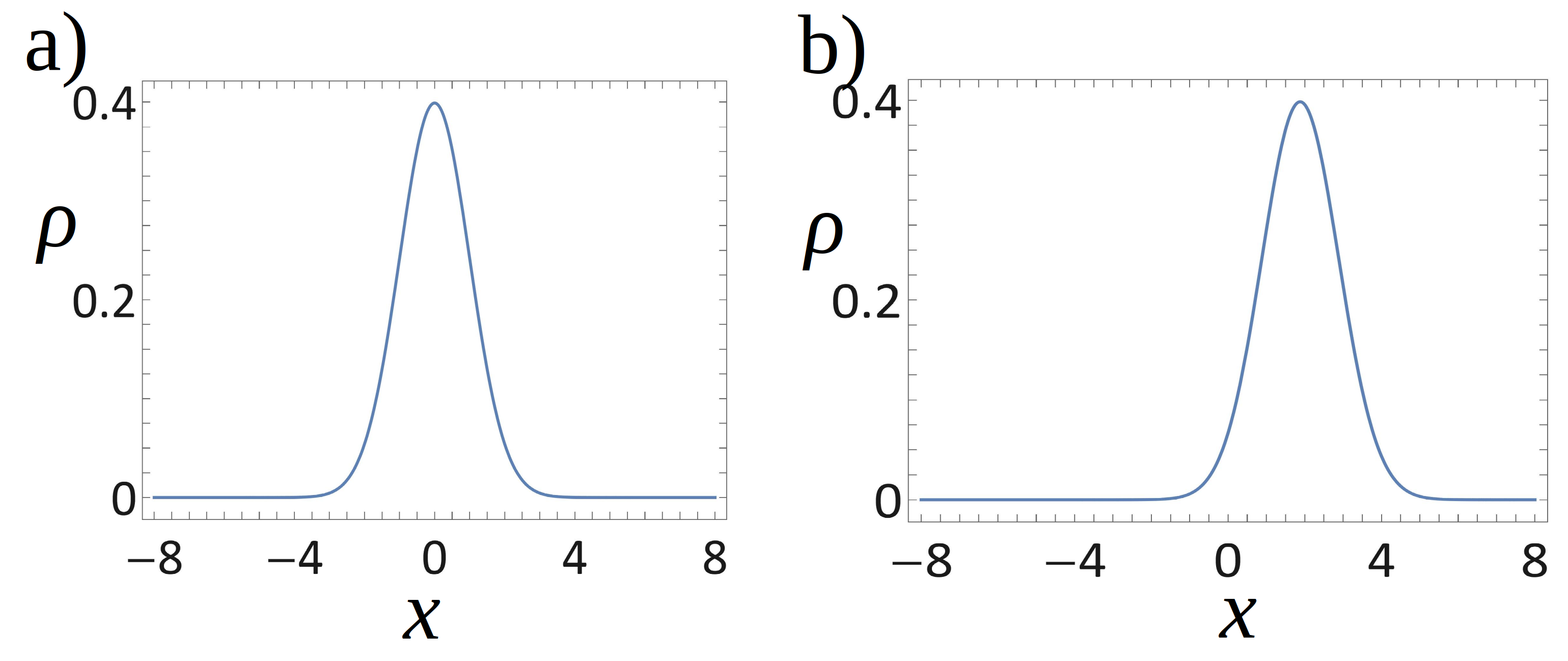}\caption{The uncoupled Klein-Gordon density of
		a spin-0 boson (a) at $t=0$ and (b) at $t=2$ (in natural units). The density at $t=0$ was chosen to be identical to the one shown in Fig. \ref{fig-ex-vel}(a). By construction, the density (\ref{RHO}) remains positive as the initial state evolves, as can be seen in panel (b).}%
	\label{fig-v-u}%
\end{figure}

\section{Defining de Broglie-Bohm trajectories in the uncoupled
representation}

\subsection{Velocity field}

We have argued above that the trouble in defining de Broglie-Bohm trajectories
from the canonical KG current comes from the fact that in the canonical
representation particles and anti-particles are mixed, so that the resulting
charge density and current results from the quantum superposition of particle
and anti-particle contributions. We propose to introduce the Bohmian velocity
field from the density and current defined from the uncoupled representation.
Using Eqs. (\ref{RHO}) and (\ref{J}), this gives%
\begin{equation}
v(t,x)=\frac{J(x,t)}{\varrho(x,t)}. \label{velu}%
\end{equation}

\begin{figure}[t]
	\centering \includegraphics[width=9cm]{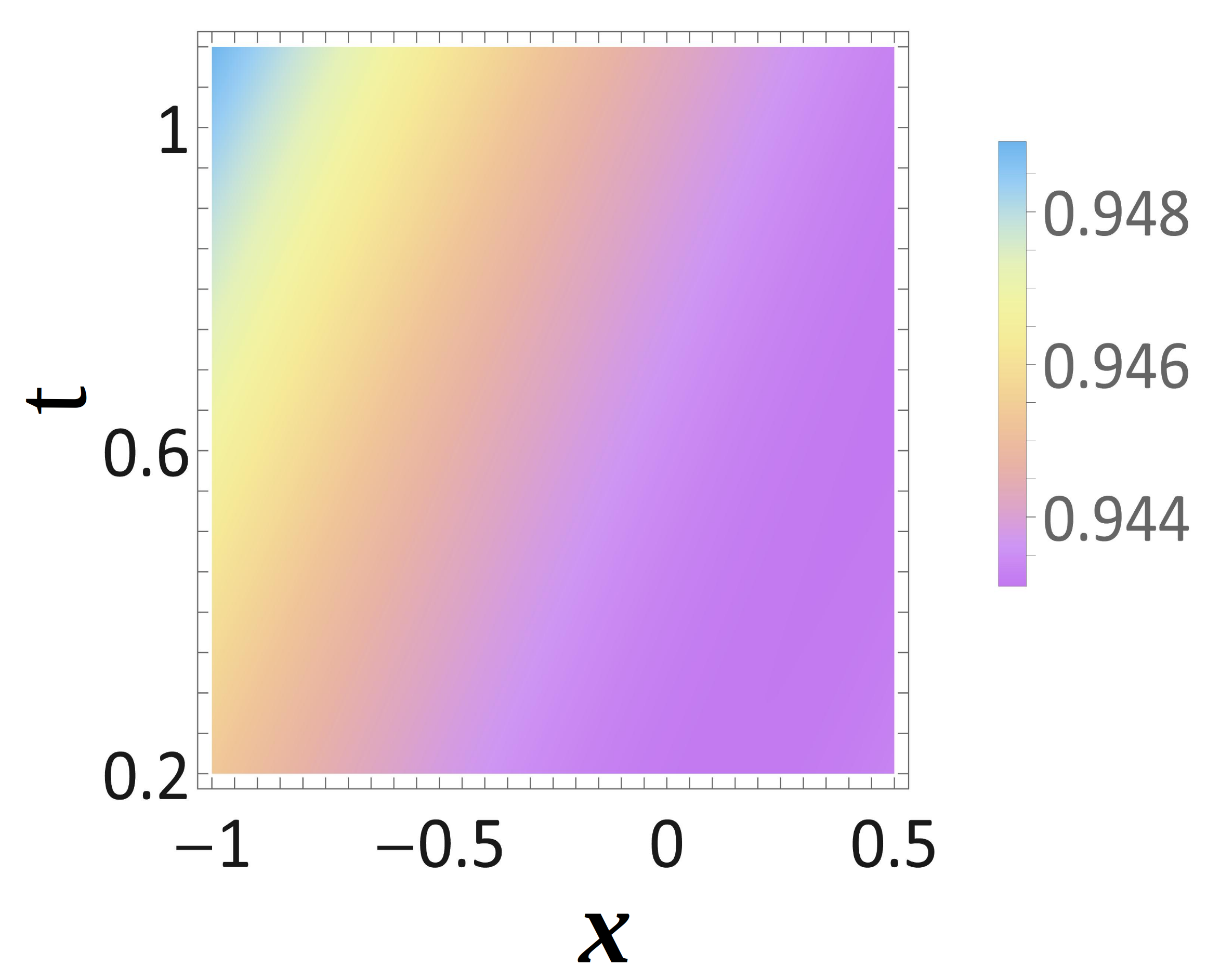}\caption{The velocity field defined by Eq. (\ref{velu}) (uncoupled Klein-Gordon velocity field) is shown (in natural units) in a given space-time region when the initial state is the one shown in Fig. \ref{fig-v-u}(a). The values are all very close to the value of the classical velocity field of a particle having momentum $p_0$ (compare with the canonical velocity field shown in Fig. \ref{fig-vel-map-c}).}%
	\label{fig-vel-map-u}%
\end{figure}

By construction, this velocity field will not suffer from the problems due to the fluctuating sign of $\rho(x,t)$ that affect the velocity
field (\ref{vel-c}) defined from the canonical KG density. As an
illustration, we reconsider the example shown in Figs. \ref{fig-ex-vel}%
-\ref{fig-vel-map-c}. The evolution of the same Gaussian leads to a density that
is everywhere positive (Fig. \ref{fig-v-u}(a)-(b)). Even though this is a
highly relativistic regime ($p_{0}=3mc$), the velocity field (\ref{velu}) is
well-behaved -- it takes values around the classical velocity (\ref{class}) of a particle 
of momentum $p_0$
and does not lead to the appearance of superluminal velocities, as pictured in
Fig. \ref{fig-vel-map-u}.

\subsection{Bohmian trajectories}

\begin{figure}[t]
	\centering \includegraphics[width=12cm]{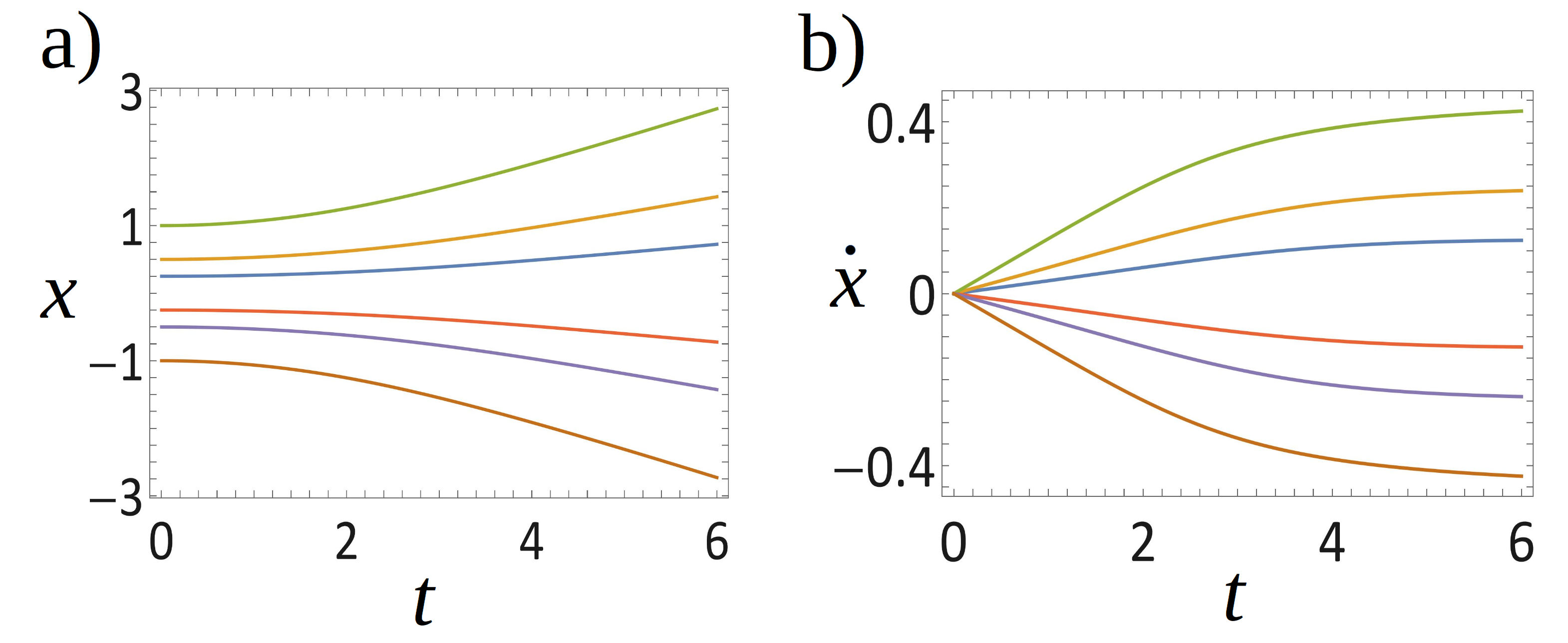}\caption{(a) Bohmian trajectories $x(t)$ for spin-0 bosons in the uncoupled Klein-Gordon representation. The  initial state is similar to the Gaussian shown in Fig. \ref{fig-v-u}(a) except for the mean momentum, taken here to be $p_0=0$ (the momentum distribution has therefore both positive and negative components). The initial position $x^0$ of each trajectory can be read off on the vertical axis. (b) Velocity $\dot{x}=dx(t)/dt$ of each Bohmian trajectory plotted in panel (a). Note that all the velocities remain sub-luminal (natural units are used, with $c=1$).}%
	\label{fig-traj-vel}%
\end{figure}

Bohmian trajectories are obtained in the usual way by solving for
$dx/dt=v[x(t),t)]$, with the initial condition $x(0)=x^0$, where $x^0$ lies
within the initial distribution $\varrho(t=0)$. An illustration is given in
Fig. \ref{fig-traj-vel}, in which trajectories and velocities obtained from
the uncoupled density and current density are shown, for an initial state
identical to the one displayed in Fig. \ref{fig-v-u}(a) except for the value
of $p_{0}$ that is taken as $p_{0}=0.$ Note that these trajectories are
similar in shape to the well-known non-relativistic de Broglie-Bohm
trajectories, and show no pathological behavior, contrary to the ones defined from the
canonical KG\ quantities.

We have not been able nevertheless to obtain strict conditions on the initial
distribution ensuring that $v(x,t)$ remains subliminal, even in the case of
free propagation. Numerical simulations starting from normalizable densities,
including with interfering distributions and the presence of nodal structures,
have all shown Bohmian trajectories with subluminal velocities, with
$J(x,t)\rightarrow0$ as $\varrho(x,t)$ vanishes. 

One particular instance for
which a superluminal velocity might appear concerns the exponentially small
propagation of the wavefunction beyond the light cone. This is a feature
obtained when a wavefunction with initial compact support is evolved with the
particle (positive energy sector) propagator only. The issues with causality
are usually disregarded, although the fundamental implications of this effect
are poorly understood. Note that this feature is neither specific to the de
Broglie Bohm formulation nor to the uncoupled representation; while the
propagator in the canonical representation is fully causal, the position
eigenfunctions are not delta functions but have a certain width of the order
of the Compton wavelength (so that a position eigenfunction lying slightly
outside the lightcone will overlap with the propagated wavefunction).

\section{Conclusion\label{sec4}}

We have seen in this paper how de Broglie-Bohm trajectories for massive particles
obeying the Klein-Gordon equation can be obtained by working with densities
and currents defined from the wavefunctions in an uncoupled representation --
a representation in which the particle (positive energy) and anti-particles
(negative energy) sectors are separated. This solves in principle the main
specific issues affecting Bohmian trajectories obtained from the canonical
representation of the KG equation, that we have attributed to the superposition of
positive and negative energy contributions. It should be emphasized that the uncoupled representation is instrumental in order to understand the physical meaning of the quantum operators
and to establish relations involving semi-classical expansions, the classical limit, and the non-relativistic limit as well.

However, additional work is needed in order to understand the dynamical
properties of the trajectories, for different initial distributions and in the
presence of potential and vector couplings. The matrix $U$ uncoupling
the particle from the anti-particle components depends on the Hamiltonian and can seldom be 
obtained in closed form 
as was the case in Eq. (\ref{FW-mat}). For Hamiltonians with arbitrary potentials one must then resort to iterative methods \cite{silenko,jentsch} in order to find the Foldy-Wouthuysen transformation. 

Most of the issues will actually
hinge on the properties of the uncoupled densities and current densities.
These quantities are not well-known, and their use remains controversial (see
e.g. the recent criticism  \cite{FW-byal} and the reply \cite{FW-reply}). The
main fundamental issue is the lack of manifest covariance of the density
\cite{rembl}: no covariant 4-current can be constructed if Eq. (\ref{RHO}) is
taken as the 0-component of such a current \cite{hoffmann}. Note that the
existence of a preferred frame, hypothesized by Bell  \cite{bell}, necessarily
appears in one form or another in relativistic Bohmian mechanics
\cite{durr,drezet}. Finally it should remembered that the first quantized
formalism breaks down when particle creation and annihilation cannot be
neglected. At this point a proper quantum field theory treatment is needed;
QFT accounts within the de Broglie-Bohm viewpoint have been proposed
\cite{jpa-qft,struyve,niko-qft}.

To sum up, we have introduced a method to define and compute Bohmian trajectories for the massive Klein-Gordon equation. As a proof of principle we have determined such trajectories for a free Hamiltonian, and checked that their properties are devoid of the serious problems that have led up to now to renounce to a trajectory interpretation for spin-0 bosons. This work hence fills a gap between the non-relativistic trajectories and the quantum field theory accounts of the de Broglie-Bohm model.


\begin{thebibliography}{99}                                                                                               %
	
	\bibitem {dB}L. De Broglie, J. Phys. Radium, 8, .225 (1927).
	
	\bibitem{bohm} D. Bohm, Phys. Rev. 85, 166 (1952) ; Phys. Rev. 85, 180 (1952).
	
		
	\bibitem {undivided}D.\ Bohm and B.\ J.\ Hiley, \textit{The Undivided
		Universe} (Routledge, London, 1993).
	

	
	\bibitem{dewd} G. Horton, C. Dewdney, and U. Ne'eman, Found. Phys. 32, 463 (2002).
	
	\bibitem {holland-PR}P. R.Holland, Phys.\ Rep. 224, 95 (1993).
	
	\bibitem {kypra}T. Kyprianidis, Phys.\ Lett.\ A 111, 111 (1985).
	
	\bibitem {nikolic08}H.\ Nikolic, Found.\ Phys. 38, 869 (2008).
	
	\bibitem {vigier-H}N. Cufaro-Petronia, C. Dewdney, P. Holland, T. Kyprianidis and
	J. P. Vigier, Phys. Lett. A  106, 368 (1984).
	
	\bibitem{dewdney2000} G. Horton, C. Dewdney and A. Nesteruk, J. Phys. A: Math. Gen. 33 7337  (2000).
	
	\bibitem {holland2019}P.\ Holland, Eur. Phys. J. Plus 134, 434 (2019).
	
	\bibitem {vollick2021} D. N. Vollick, Can. J. Phys. 99, 100 (2021).
	
	\bibitem {FW}L. L. Foldy and S. A. Wouthuysen, Phys. Rev. 78, 29 (1950).
	
	
	\bibitem {case}K. M. Case, Phys. Rev. 95, 1323 (1954).
	
	\bibitem {FV}H.\ Feschbach and F.\ Villars, Rev.\ Mod.\ Phys.\ 30, 24 (1958).
	
	\bibitem {silenko}L. Zou, P. Zhang, and A. J. Silenko, Phys. Rev. A 101,
	032117 (2020).
	\bibitem {hoffmann}S. E Hoffmann,  J. Phys. A: Math. Theor. 52 225301 (2019).
	
	\bibitem{leavens}C. R. Leavens, Found. Phys. 35, 469 (2005). 
	
	\bibitem{WVreview}A. Matzkin, Found. Phys. 49, 298 (2019),
	
	\bibitem {greiner}W.\ Greiner, \emph{Relativistic Quantum Mechanics}
	(Springer, Berlin, 1996).
	
	\bibitem {mosta}A.\ Mostafazadeh, Class. Quantum Grav. 20 155 (2003).
	
	\bibitem {mak}M. Alkhateeb, X. Gutiérrez de la Cal, M. Pons, D. Sokolovski,
	and A. Matzkin, Phys. Rev. A 103, 042203 (2021).
	
	\bibitem{holland-book}P. R. Holland, \emph{The Quantum Theory of Motion} (Cambridge University Press, Cambridge, England, 1993), Sec. 12.1.
	
	\bibitem{debroglie-KGsup}L. de Broglie, \textit{Non-linear wave mechanics} (Elsevier, Amsterdam, 1960), Ch. 9, Sec. 2. [originally published as L. de Broglie, \emph{Une interprétation causale et non linéaire de la Mécanique ondulatoire: la théorie de la double solution} (Gauthier-Villars, Paris, 1956)].
	
	\bibitem {nikolic04}H.\ Nikolic, Found.\ Phys. Lett. 17, 363 (2004).
	
	\bibitem {pitfall}A.\ Matzkin and V.\ Nurock, Studies in Hist. and Phil.
	of Science B 39, 17 (2008).
	
	\bibitem {billiard}A.\ Matzkin, Found. Phys. 39, 903 (2009).
	
	\bibitem {NW}T. D. Newton and E. P. Wigner, Rev. Mod. Phys. 21, 400 (1949).
	
	\bibitem {costella}J. P. Costella and B. H. J. McKellar, Am. J. Phys. 63 1119 (1995).
	
		\bibitem {kowalski}K. Kowalski and J. Rembielinski, Phys. Rev. A 84, 012108 (2011).
	
	\bibitem{jentsch}A. Wienczek, C. Moore and U. D. Jentschura, Phys. Rev. A 106, 012816 (2022).
	
	\bibitem{FW-byal}I. Bialynicki-Birula and Z. Bialynicka-Birula, Phys. Rev. Lett. 122, 159301 (2019).
	
	
	\bibitem{FW-reply} A. J. Silenko, P. Zhang, and L. Zou, Phys. Rev. Lett. 122, 159302  (2019).
	
	
	\bibitem {rembl}J. Rembielinski and K. A. Smolinski, EPL 88 10005 (2009).
	
	
	\bibitem {bell}J.\ S\ Bell, 
	``Quantum mechanics for cosmologists'', reprinted in
	J.\ S\ Bell, \emph{Speakable and unspeakable in quantum mechanics}, 2nd
	edition\emph{ }(Cambridge University Press, Cambridge, England, 2004), Ch.\ 15.
	
	\bibitem {durr}D. Durr, S. Goldstein, T. Norsen, W. Struyve and N. Zanghi,
	Proc. R. Soc. A.470 20130699 (2014).
	
	\bibitem {drezet}A. Drezet, Found Phys 49, 1166 (2019).
		\bibitem {jpa-qft}D. Durr, S. Goldstein, R. Tumulka and N. Zanghi,  J. Phys.
	A: Math. Gen. 36 4143 (2003).
	
	\bibitem {struyve}W. Struyve, Rep. Prog. Phys. 73 106001 (2010).
	
	\bibitem {niko-qft}H.\ Nikolic, Int. J. Mod. Phys. A 25, 1477 (2010).
	
	
\end{thebibliography}
\end{document}